\documentclass[%
aps,%
prb,%
reprint,%
amsmath,amssymb,%
floatfix,%
nofootinbib,%
superscriptaddress%
]{revtex4-2}

\usepackage[T1]{fontenc}
\usepackage[latin9]{inputenc}
\usepackage[english]{babel}
\usepackage{mathrsfs}
\usepackage{bbm}
\usepackage{graphicx}
\usepackage{xcolor}
\usepackage{physics}
\usepackage{bm}
\usepackage[export]{adjustbox}

\usepackage[colorlinks,
            linkcolor=blue,    
            citecolor=blue,  
            urlcolor=blue,
 	        bookmarks=true,        
	        bookmarksopen=true,    
	        bookmarksnumbered=true,
]{hyperref}

\renewcommand*\d{\mathop{}\!\mathrm{d}}

\newcommand{\av}[1]{\left\langle#1\right\rangle}

\newcommand{\conj}[1]{{#1}^*}
\newcommand{\id}{\mathbbm{1}}

\newcommand{\sT}{\mathscr{T}}
\newcommand{\sP}{\mathcal{P}}
\newcommand{\sQ}{\mathcal{Q}}
\renewcommand{\tt}{\mathbf{t}}

\newcommand{\RH}{\check{\mathcal R}}
\newcommand{\RHp}{\mathcal R}

\begin{document}

\title{Integrable nonunitary open quantum circuits}

\author{Lucas  S\'a}
\email{lucas.seara.sa@tecnico.ulisboa.pt}

\affiliation{CeFEMA, Instituto Superior T\'ecnico, Universidade de Lisboa, Av.\ Rovisco Pais, 1049-001 Lisboa, Portugal}

\author{Pedro Ribeiro}
\email{ribeiro.pedro@tecnico.ulisboa.pt}

\affiliation{CeFEMA, Instituto Superior T\'ecnico, Universidade de Lisboa, Av.\ Rovisco Pais, 1049-001 Lisboa, Portugal}
\affiliation{Beijing Computational Science Research Center, Beijing 100193, China}

\author{Toma\v z Prosen}
\email{tomaz.prosen@fmf.uni-lj.si}

\affiliation{Department of Physics, Faculty of Mathematics and Physics, University of Ljubljana, Jadranska 19, SI-1000 Ljubljana, Slovenia}

\begin{abstract}
We explicitly construct an integrable and strongly interacting dissipative quantum circuit via a trotterization of the Hubbard model with imaginary interaction strength.  To prove integrability, we build an inhomogeneous transfer matrix, from which conserved superoperator charges can be derived, in particular, the circuit's dynamical generator. After showing the trace preservation and complete positivity of local maps, we reinterpret them as the Kraus representation of the local dynamics of free fermions with single-site dephasing. The integrability of the map is broken by adding interactions to the local coherent dynamics or by removing the dephasing. In particular, even circuits built from convex combinations of local free-fermion unitaries are nonintegrable. Moreover, the construction allows us to explicitly build circuits belonging to different non-Hermitian symmetry classes, which are characterized by the behavior under transposition instead of complex conjugation. We confirm all our analytical results by using complex spacing ratios to examine the spectral statistics of the dissipative circuits.
\end{abstract}

\maketitle

\section{Introduction}%
Integrability is a fascinating field of mathematical physics. It provides exact solutions to dynamics and equilibrium in very diverse contexts, ranging from deterministic (i) classical~\cite{faddeevtakhtajan,bernard} and (ii) quantum~\cite{faddeev,korepin} many-body Hamiltonian dynamics, to classical stochastic systems, (iii) {\em in}~\cite{baxter1982}, and (iv) {\em out}~\cite{derrida} of equilibrium.
Although the Liouville-Arnold~\cite{arnold} (i), Bethe-ansatz~\cite{bethe,gaudin} (ii), and 
Onsager~\cite{onsager} (iii) threads of integrability were initially developed independently, they were beautifully united within the techniques of (quantum) inverse scattering~\cite{gardner,korepin,faddeev} and the celebrated Yang-Baxter equation~\cite{baxter1982,cnyang}.

Later, quantum inverse scattering methods (a.k.a.\ algebraic Bethe ansatz) found their way to the exact solution (diagonalization) of classical stochastic systems---many-body Markov chains, such as simple exclusion processes~\cite{kirone}. More recently, related new techniques have been developed for the exact solution of open integrable quantum many-body systems, specifically, by extending the algebraic Bethe ansatz to noncompact (nonunitary) auxiliary spaces~\cite{prosen2015REVIEW} and by providing an exact mapping between Liouvillians of open many-body systems and Bethe-ansatz integrable systems on (thermofield) doubled Hilbert spaces~\cite{medvedyeva2016,essler2020}.

Very recently, (local) quantum circuits have become an important paradigm of nonequilibrium many-body physics, in particular, due to their simulability by emerging quantum computing facilities, where they provide a natural platform for the demonstration of quantum supremacy~\cite{google}. Moreover, (open) quantum circuits with local projective measurements have been shown to host an exciting new physics paradigm of measurement-induced phase transitions~\cite{chan2019,skinner2019,li2019,altman,ludwig}.

The natural and significant question arises, if integrability methods can be extended to such a paradigm. The results on integrable trotterizations of integrable quantum spin chains~\cite{vanicat2018} and classical stochastic parallel update exclusion processes~\cite{vanicat2018b} give very encouraging hints.

In this paper, we make a key step in this direction by constructing an integrable open (nonunitary) local quantum circuit. We show that Shastry's $\check{R}$-matrix~\cite{shastry1986a,shastry1986b,shastry1988,maassarani1998,essler2005}, the essential integrability concept of the one-dimensional Fermi-Hubbard model, can be interpreted as a completely positive (CP) trace-preserving (TP) map over a pair of qubits (spins 1/2) after a suitable analytic continuation of the interaction and spectral parameters. Our CPTP map represents a convex combination of two coherent (unitary) symmetric nearest-neighbor-hopping (XX) processes, one of them composed with local dephasing. By virtue of the Yang-Baxter equation, we then show the existence of a commuting transfer matrix for the brickwork quantum circuit built from such CPTP maps, generating a family of local superoperators commuting with the dynamical map. Integrability of the Floquet dynamics is also demonstrated empirically by studying spectral statistics (complex spacing ratios~\cite{sa2019CSR}), whose sensitivity to integrability breaking is shown by studying two alternative families of local open quantum circuits.

The rest of the paper is organized as follows. In Sec.~\ref{sec:circuit} we define the dissipative Hubbard circuit and describe in detail its elementary local gates. Next, we prove that the circuit is indeed integrable and CPTP in Secs.~\ref{sec:proof_integrability} and \ref{sec:proof_CPTP}, respectively. The subsequent three sections focus on the physical content of our circuit: We address its Kraus representation in Sec.~\ref{sec:Kraus_representation}, integrability-breaking regimes in Sec.~\ref{sec:integrability_breaking}, and its symmetries in Sec.~\ref{sec:symmetries}. Finally, we present numerical evidence corroborating all our results in Sec.~\ref{sec:numerics} before drawing conclusions and summarizing our findings in Sec.~\ref{sec:conclusions}. Three Appendices present some additional details.

\section{The dissipative Hubbard circuit}
\label{sec:circuit}
We consider a spin-$1/2$ chain of even size $L$ with periodic boundary conditions. The density matrix $\rho$ of the system evolves under the action of the discrete-time quantum channel $\Psi$, $\rho(t+1)=\Psi[\rho(t)]$---a linear map over the $4^L$-dimensional state vector $\rho$---that we choose to be of the brickwork circuit form:
\begin{equation}\label{eq:Psi_def}
\begin{split}
\Psi&=
\left(
\prod_{j=1}^{L/2}
\RH_{2j,2j+1}
\right)
\left(
\prod_{j=1}^{L/2}\RH_{2j-1,2j}
\right)
\\
&=\includegraphics[width=0.82\columnwidth,valign=c]{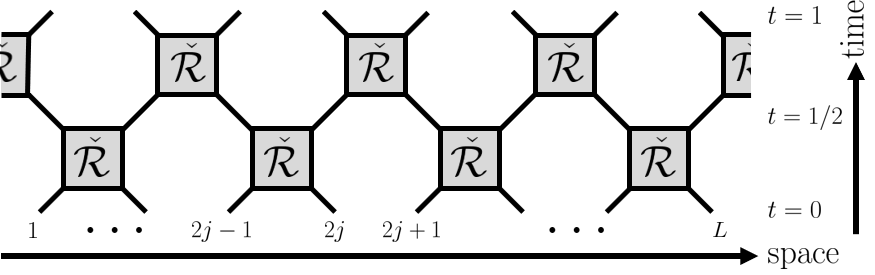}
\end{split}
\end{equation}
Here, $\RH_{kl}$ is the Hubbard $\check{R}$-matrix nontrivially acting on sites $k$ and $l$. Each wire in Eq.~(\ref{eq:Psi_def}) carries a four-dimensional operator Hilbert space and $\RH$ acts as a two-site ($16\times16$) elementary gate (grey box). One time-step consists of two rows of the circuit---in the second of which the elementary gates are shifted by one site. Accordingly, Eq.~(\ref{eq:Psi_def}) can also be written as
\begin{equation}
    \Psi=\mathbb{T}^\dagger\Phi\mathbb{T}\Phi,
\end{equation} 
where $\Phi=\RH^{\otimes L/2}$ corresponds to a single row of the circuit. We introduced the one-site translation operator $\mathbb{T}$, defined by its action on the computational operator-basis, $\mathbb{T}\ket{e_1,e_2,\dots,e_L}=\ket{e_L,e_1,\dots,e_{L-1}}$, where 
indices $e_j\in\{0,1,2,3\}$ label four possible spin-1/2 operators at site $j$.
While each row $\Phi$ of the circuit is factorizable into two-site elementary gates, the checkerboard pattern renders the full circuit $\Psi$ interacting. Because the same gate $\RH$ is applied throughout space and time, the repeated action of $\Psi$ leads to, in general, nonunitary translationally-invariant Floquet dynamics.

After a Jordan-Wigner transformation, the  Hubbard model can be understood as a spin ladder formed of a pair of XX models (corresponding to up- and down-spin fermions or to the \emph{bra} and the \emph{ket} of the density matrix~\cite{medvedyeva2016} in our nonunitary formulation) coupled by the Hubbard interaction along the rungs. Thus, we start with the (two-site) spin-$1/2$ XX $\check{R}$-matrix,
\begin{equation}\label{eq:Rch_general6v}
    \check{R}
    =\frac{1}{a}
    \begin{pmatrix}
    a & 0 & 0 & 0 \\
    0 & c & -i b & 0 \\
    0 & i b & c & 0 \\
    0 & 0 & 0 & a
    \end{pmatrix},
\end{equation}
which admits a simple trigonometric parametrization:
\begin{equation}\label{eq:RchXX}
    a=\cos\lambda,\quad
    b=\sin\lambda,\quad
    c=1.
\end{equation}
$\check{R}=\check{R}(\lambda)$ is real orthogonal for imaginary spectral parameter $\lambda\in i\mathbb R$.\footnote{We have introduced the factors of $\pm i$ multiplying $b$ (corresponding to the choice $x=-i$ in Eq.~(12.93) of Ref.~\cite{essler2005}) to compensate for the imaginary spectral parameter (since $\sin i\lambda=i\sinh\lambda$). While these factors could be removed by a trivial similarity transformation, this choice will prove convenient below.}
We introduce a basis $\{e_{\alpha}^{\beta}\}$ of $2\times2$ matrices such that the only nonzero entry (equal to 1) of $e^{\beta}_{\alpha}$ is in row $\alpha$ and column $\beta$. We then consider the action of $\check{R}$ on two copies of the system (corresponding to ket ($\uparrow$) and bra ($\downarrow$) of the vectorized density matrix $\rho=\sum_{mn}\rho_{mn} \ket{m}\bra{n}\mapsto \ket{\rho}=
\sum_{mn}\rho_{mn}\ket{m}\otimes \ket{n}^*$),
\begin{equation}  \label{eq:rupdown}
\begin{split}
\check{r}_{\uparrow}(\lambda)=\check{R}^{\alpha\gamma}_{\beta\delta}(\lambda)\,e^{\beta}_{\alpha}\otimes \id_2\otimes e^{\delta}_{\gamma}\otimes \id_2,\\
\check{r}_{\downarrow}(\lambda)=\check{R}^{\alpha\gamma}_{\beta\delta}(\lambda)\,\id_2\otimes e^{\beta}_{\alpha}\otimes \id_2\otimes e^{\delta}_{\gamma},
\end{split}
\end{equation}
where $\id_d$ is the $d\times d$ identity matrix. Summation over repeated indices is assumed throughout. 

In terms of the XX $\check{R}$-matrices, the Hubbard $\check{R}$-matrix reads (choosing the appropriate gauge)~\cite{essler2005}
\begin{equation}\label{eq:RcheckHubb}
\RH(\lambda,\mu)
=\beta\, \check{r}(\lambda-\mu)
+\alpha\, \check{r}(\lambda+\mu)
\left(\sigma^{z}\otimes\sigma^{z}\otimes 
\id_4\right),
\end{equation}  
where $\check{r}(\lambda)=\check{r}_{\uparrow}(\lambda)\check{r}_{\downarrow}(\lambda)$ and $\sigma^{x,y,z}$ denote the standard Pauli matrices. The two prefactors $\alpha\equiv\alpha(\lambda,\mu,u)$ and $\beta\equiv\beta(\lambda,\mu,u)$ depend on two independent spectral parameters $\lambda$ and $\mu$ and on the Hubbard interaction strength $u$. The Hubbard $\check{R}$-matrix---which is not of difference form---satisfies the Yang-Baxter equation,
\begin{equation}\label{eq:YBE}
\begin{split}
&\left(\id_4\otimes\RH(\lambda,\mu)\right)
\left(\RH(\lambda,\nu)\otimes\id_4\right)
\left(\id_4\otimes\RH(\mu,\nu)\right)=\\
=&\left(\RH(\mu,\nu)\otimes\id_4\right)
\left(\id_4\otimes\RH(\lambda,\nu)\right)
\left(\RH(\lambda,\mu)\otimes \id_4\right)
,
\end{split}
\end{equation}
if the ratio $\alpha/\beta$ is fixed as
\begin{equation}
\frac{\alpha}{\beta}=\frac{\cos(\lambda+\mu)\sinh(h-\ell)}{\cos(\lambda-\mu)\cosh(h-\ell)},
\end{equation}
where $h$ and $\ell$ are implicitly defined in terms of $\lambda$, $\mu$, and $u$ through $\sinh(2h)/\sin(2\lambda)=\sinh(2\ell)/\sin(2\mu)=u$. Finally, by choosing
\begin{equation}
\beta=
\frac{\cos(\lambda-\mu)\cosh(h-\ell)}{\cos(\lambda-\mu)\cosh(h-\ell)+\cos(\lambda+\mu)\sinh(h-\ell)},
\end{equation}
we have $\alpha+\beta=1$. Furthermore, with this choice of $\beta$, $\RH$ satisfies the unitarity condition:
\begin{equation}\label{eq:unitarity_condition}
\RH(\lambda,\mu)\RH(\mu,\lambda)=\id_{16}.
\end{equation}

\section{Proof of the integrability of the Hubbard circuit}
\label{sec:proof_integrability}
By construction, the Hubbard circuit $\Psi$~(\ref{eq:Psi_def}) is integrable. Indeed, since $\RH$ satisfies the (braid) Yang-Baxter equation~(\ref{eq:YBE}), there exists a one-parameter family $\tt(\omega)$ of transfer matrices in involution, i.e.,  $\comm{\tt(\omega_1)}{\tt(\omega_2)}=0$ for all $\omega_1,\omega_2$. After introducing an auxiliary space, labeled $a$, identical to the (local) four-dimensional physical Hilbert space, the transfer matrix is expressed as the partial trace of the monodromy matrix, $\tt(\omega)=\Tr_a\sT_a(\omega)$, with
\begin{equation}\label{eq:monodromy_matrix}
\sT_a(\omega)=\prod_{1\leq j\leq L}^{\leftarrow}\RHp_{aj}\left(
\omega,\frac{\lambda+\mu}{2}-(-1)^j\frac{\lambda-\mu}{2}
\right),
\end{equation}
where $\RHp=\sP \RH$, $\sP$ is a $16\times 16$ permutation matrix defined by $\sP \left( \ket{\rho_1} \otimes \ket{\rho_2} \right) = \ket{\rho_2}\otimes \ket{\rho_1}$, and the symbol {\scriptsize $\displaystyle{\prod^{\leftarrow}}_j$} indicates an ordered product with decreasing index~$j$. The monodromy matrix $\sT_a$ is inhomogeneous (staggered) to account for the checkerboard pattern of the quantum circuit. Evaluating the monodromy matrix~(\ref{eq:monodromy_matrix}) at the two special (a.k.a.\ shift) points $\omega=\lambda$ and $\omega=\mu$, the Floquet propagator $\Psi$, defined by Eq.~(\ref{eq:Psi_def}), can be written as
\begin{equation}\label{eq:Psi_TransferMatrix}
\Psi=\tt(\mu)^{-1}\tt(\lambda).
\end{equation} 
To verify this claim, we start by computing the monodromy matrix at $\omega=\lambda$. It reads as
\begin{equation}
\begin{split}
\sT_a(\lambda)\,
&=\prod_{1\leq j\leq L/2}^{\leftarrow}
\RHp_{a,2j}(\lambda,\mu)\sP_{a,2j-1}\\
&=\prod_{1\leq j\leq L/2}^{\leftarrow}
\sP_{a,2j}\RH_{a,2j}(\lambda,\mu)\sP_{a,2j-1}\\
&=\prod_{1\leq j\leq L/2}^{\leftarrow}
\sP_{a,2j}\sP_{a,2j-1}
\RH_{2j-1,2j}(\lambda,\mu),
\end{split}
\end{equation}
where we have used the identities $\RHp(\lambda,\lambda)=\RHp(\mu,\mu)=\sP$ and $\RH_{al}\sP_{ak}=\sP_{ak}\RH_{kl}$. Because all $\sP$ and $\RH$ operators commute with each other when acting on different Hilbert spaces (i.e., when they have no subscript indices in common), taking the trace over the auxiliary space yields
\begin{equation}
\tt(\lambda)=\Tr_a\left[
\prod_{1\leq j\leq L}^{\leftarrow}\sP_{aj}
\right]\prod_{j=1}^{L/2}\RH_{2j-1,2j}(\lambda,\mu).
\end{equation} 
To evaluate the remaining trace, we use $\sP_{al}\sP_{ak}=\sP_{ak}\sP_{kl}$ to permute $\sP_{a1}$ over all the other $\sP_{aj}$ and then use $\Tr_a \sP_{a1}=\id_4$. The transfer matrix finally reads as
\begin{equation}
\tt(\lambda)=\left(
\prod_{2\leq j\leq L}^{\leftarrow}\sP_{1j}
\right)
\prod_{j=1}^{L/2}\RH_{2j-1,2j}(\lambda,\mu).
\end{equation}
The computation for $\omega=\mu$ proceeds similarly and results in the expression
\begin{equation}
\sT_a(\mu)
=\left(
\prod_{1\leq j\leq L}^{\leftarrow} \sP_{aj}
\right)
\RH_{a1}(\mu,\lambda)
\prod_{j=1}^{L/2-1}\RH_{2j,2j+1}(\mu,\lambda).
\end{equation}
To evaluate the trace over the auxiliary space, we first cycle $\RH_{a1}$ to the left of the product of permutations, permute it over $\sP_{aL}$ to obtain $\RH_{L1}$, take it out of the trace, and, at last, evaluate the resulting trace of permutations as above. Finally, imposing periodic boundary conditions (i.e., identifying $j=L+1$ with $j=1$) and using the unitarity condition (\ref{eq:unitarity_condition}), the transfer matrix at $\omega=\mu$ is given by 
\begin{equation}
\tt(\mu)=\left(
\prod_{2\leq j\leq L}^{\leftarrow}\sP_{1j}
\right)
\prod_{j=1}^{L/2}\left(\RH_{2j,2j+1}(\lambda,\mu)\right)^{-1}.
\end{equation}
It is now evident that the dynamical generator (\ref{eq:Psi_def}) can be written as in Eq.~(\ref{eq:Psi_TransferMatrix}):
\begin{equation}
\begin{split}
\Psi&=\tt(\mu)^{-1}\tt(\lambda)\\
&=
\left(
\prod_{j=1}^{L/2}\RH_{2j,2j+1}(\lambda,\mu)
\right)
\left(
\prod_{j=1}^{L/2}\RH_{2j-1,2j}(\lambda,\mu)
\right).
\end{split}
\end{equation}

The involution property of the transfer matrix implies the integrability of the circuit since $\Psi$ commutes with $\tt(\omega)$ for all $\omega$ and, in particular, with the two infinite sets of local superoperator charges generated from $\tt(\omega)$ by logarithmic differentiation:
\begin{equation}
\sQ_n^{(1)}=\frac{\d^n}{\d\omega^n}\log \tt(\omega)\bigg|_{\omega=\lambda},\ 
\sQ_n^{(2)}=\frac{\d^n}{\d\omega^n}\log \tt(\omega)\bigg|_{\omega=\mu}.
\end{equation}

\section{The Hubbard \texorpdfstring{$\check{R}$}{R}-matrix as a local CPTP map}
\label{sec:proof_CPTP}
Having proved that the circuit is integrable, it remains to be shown that it describes proper open quantum dynamics, i.e., that it is a CPTP map. It suffices to show this for the elementary gates $\RH$. Indeed, choosing $\lambda,\mu,u \in i\mathbb{R}$ (purely imaginary interaction), then $\alpha,\beta\in\mathbb{R}$ and $\RH$ becomes a bistochastic quantum map~\cite{bengtsson2017,bruzda2009} (i.e., a unital CPTP map). To check this result, we first reshuffle the indices of $\RH$ to obtain the dynamical Choi matrix~$D$~\cite{bengtsson2017}, such that 
$D^{\alpha\gamma\varepsilon\eta}_{\beta\delta\zeta\theta}=\RH^{\alpha\beta\varepsilon\zeta}_{\gamma\delta\eta\theta}$.
Due to the channel-state duality~\cite{jamiolkowski1972,choi1975}, the map $\RH$ is CP if $D$ is non-negative; it is TP if the partial trace of $D$ over the first copy of the system is the identity; and it is unital if the partial trace over the second copy of the system is the identity. The TP and unitary conditions can be written as
\begin{subequations}
\begin{align}\label{eq:condition_TP}
&D^{\alpha\gamma\varepsilon\eta}_{\alpha\delta\varepsilon\theta}=
\RH^{\alpha\alpha\varepsilon\varepsilon}_{\gamma\delta\eta\theta}
=\delta^{\gamma}_{\delta}\delta^{\eta}_{\theta},
\\\label{eq:condition_unital}
&D^{\alpha\gamma\varepsilon\eta}_{\beta\gamma\zeta\eta}=
\RH^{\alpha\beta\varepsilon\zeta}_{\gamma\gamma\eta\eta}
=\delta^{\alpha}_{\beta}\delta^{\varepsilon}_{\zeta},
\end{align}
\end{subequations}
respectively. To see that Eq.~(\ref{eq:condition_TP}) holds, we write out the components of the Choi matrix using Eq.~(\ref{eq:RcheckHubb}),
\begin{equation}
\begin{split}
D^{\alpha\gamma\varepsilon\eta}_{\beta\delta\zeta\theta}(\lambda,&\mu)
=\beta\, \check{R}^{\alpha\varepsilon}_{\gamma\eta}(\lambda-\mu)\check{R}^{\beta\zeta}_{\delta\theta}(\lambda-\mu)\\
&+\alpha\, \check{R}^{\alpha\varepsilon}_{\iota\eta}(\lambda+\mu)\check{R}^{\beta\zeta}_{\kappa\theta}(\lambda+\mu)\left(\sigma^z\right)^{\iota}_{\gamma}\left(\sigma^z\right)^{\kappa}_{\delta},
\end{split}
\end{equation}
compute the trace of Eq.~(\ref{eq:condition_TP}),
\begin{equation}\label{eq:proof_TP}
\begin{split}
    D^{\alpha\gamma\varepsilon\eta}_{\alpha\delta\varepsilon\theta}
    &=\left(\check{R}^\dagger \check{R}\right)^{\gamma\eta}_{\delta\theta}
    \times \begin{cases}
    \beta+\alpha\quad\text{if }\gamma=\delta\\
    \beta-\alpha\quad\text{if }\gamma\neq\delta
    \end{cases}\\
    &=(\beta+\alpha)\delta_\delta^\gamma\delta_\theta^\eta
    =\delta_\delta^\gamma\delta_\theta^\eta,
\end{split}
\end{equation}
and find that the map is indeed TP. In Eq.~(\ref{eq:proof_TP}), the three equalities hold because (i) $\check{R}$ admits a real representation, (ii) it is unitary, and (iii) we have fixed $\alpha+\beta=1$, respectively. Similarly, a computation starting from Eq.~(\ref{eq:condition_unital}) leads to a term proportional to $\check{R}\check{R}^\dagger$, which again is nonvanishing only when the prefactor is $\beta+\alpha=1$, and the map is therefore unital. Finally, since we have a rank-two map, of the sixteen eigenvalues of $D$ fourteen are zero and the remaining two are explicitly found to be $4\alpha>0$ and $4\beta>0$.\footnote{ 
The analytic continuation of $\lambda$, $\mu$, and $u$ to the imaginary axes can always be chosen to render $\alpha$ and $\beta$ positive.}
The Choi matrix is therefore non-negative and the map is CP. We have thus shown that a suitable analytic continuation of $\RH$ is a unital CPTP map.

\section{Kraus representation}
\label{sec:Kraus_representation}
The previous result implies that $\RH$ can be written in the Kraus form~\cite{kraus1983,nielsen2002,bengtsson2017}. 
Abandoning the formal identification with the Fermi-Hubbard model, we identify, by reordering tensor factors, 
\begin{equation}
    \phi^{\alpha\gamma\varepsilon\eta}_{\beta\delta\zeta\theta}(q_+,q_-,p)\equiv\RH^{\alpha\varepsilon\gamma\eta}_{\beta\zeta\delta\theta}(\lambda,\mu,u)
\end{equation}
with a vectorized quantum map parametrized by three independent real parameters: the coherent hopping strengths $q_\pm\equiv-i(\lambda\pm\mu)\in\mathbb{R}$ and the relative weight of the channels $p\equiv\alpha\in[0,1]$. 
Swapping the second and third tensor-product factors in Eq.~(\ref{eq:RcheckHubb}), $\phi$ can be written in the vectorized Kraus representation,
\begin{equation}\label{eq:Kraus_rep}
\phi(q_+,q_-,p)=K_-\otimes \conj{K}_- + K_+\otimes\conj{K}_+, 
\end{equation} 
acting on (local two-site) states as $\phi[\rho]=K_-\rho K_-^\dagger+K_+\rho K_+^\dagger$,
with Kraus operators
\begin{equation}\label{eq:Kraus_2channel}
K_-=\sqrt{1-p}\,\check{R}(i q_-),\;\;
K_+=\sqrt{p}\,\check{R}(i q_+)
\left(\sigma^z\otimes \id_2\right).
\end{equation}

We see that the Kraus map of Eqs.~(\ref{eq:Kraus_rep}) and (\ref{eq:Kraus_2channel}), which we dub the Hubbard-Kraus map, describes the discrete-time dynamics of free fermions (after undoing the Jordan-Wigner transformation) subjected to local dephasing.\footnote{
Note that, the circuit $\Psi$ does \emph{not} describe a dissipative Hubbard model. $\RH$ is used as a mathematical device to build an integrable circuit which, \textit{a priori}, is unrelated to the original model. For a recent study of an exactly-solvable dissipative Hubbard model, see Ref.~\cite{nakagawa2020}.
} 
Indeed, after a suitable change of basis, the $\check{R}$-matrix~(\ref{eq:Rch_general6v}) with parametrization~(\ref{eq:RchXX}) can be written as $\check{R}(iq_\pm)=\exp\{i\,\mathrm{gd}(q_\pm)H_{\mathrm{XX}}\}$, where $ H_\mathrm{XX}=\left(\sigma^x\otimes\sigma^x+\sigma^y\otimes\sigma^y\right)/2$ is the XX-chain Hamiltonian and $\mathrm{gd}(q)=\int_0^{q}\d x/\cosh{x}$ is the Gudermannian function. In the Trotter limit, $q_\pm\to0$, $\mathrm{gd}(q_\pm)\to q_\pm$, and the quantum map~(\ref{eq:Kraus_rep}) describes the quantum stochastic process in which, at each (discrete) half-time-step, with probability $1-p$, a fermion hops from the first to the second site (or vice-versa) with amplitude $q_-$; or, with probability $p$, it hops with amplitude $q_+$; in the latter case, it is also subject to dephasing when at the first site. We again emphasize that only the local Kraus maps describe free dynamics (with dephasing), as the checkerboard pattern of the circuit makes the full circuit strongly interacting.

\section{Breaking integrability}
\label{sec:integrability_breaking}
A natural way of breaking the integrability of the circuit is by adding interactions to the \emph{local} coherent processes. This can be achieved by replacing, in Eq.~(\ref{eq:Kraus_2channel}), the XX $\check{R}$-matrices by more general XXZ (six-vertex) $\check{R}$-matrices~\cite{baxter1982,samaj2013}, which have the same form of Eq.~(\ref{eq:Rch_general6v}) but admit a two-parameter trigonometric parametrization,
\begin{equation}\label{eq:RchXXZ}
    a=\sin(\lambda+\gamma),\quad
    b=\sin\lambda,\quad
    c=\sin\gamma,
\end{equation}
where $\gamma\in(-\pi,\pi]$ is related to the anisotropy parameter of the XXZ chain and, as before, $\lambda\in i\mathbb{R}$. The XX $\check{R}$-matrix~(\ref{eq:RchXX}) follows from Eq.~(\ref{eq:RchXXZ}) upon setting $\gamma=\pi/2$. The resulting Kraus operators~(\ref{eq:Kraus_2channel}) have five independent real parameters $p$, $q_\pm$, $\gamma_\pm$ and the extensive dephasing-XXZ circuit is built from them exactly as before. Note that integrability is broken because the $\RH$-matrix obtained this way from Eqs.~(\ref{eq:rupdown})~and~(\ref{eq:RcheckHubb}) no longer satisfies the Yang-Baxter equation~(\ref{eq:YBE}).

Furthermore, while one might be tempted to conjecture the integrability of the quantum map~(\ref{eq:Kraus_2channel}) for general $\check{R}$-matrices at a free-fermion point~\cite{fan1970} (i.e., satisfying $a^2=c^2-b^2$), this turns out not to be correct. One such model is obtained from the Hubbard-Kraus map by removing dephasing from the second channel. The resulting quantum map, still of the form~(\ref{eq:Kraus_rep}), is a convex combination of two unitary free Kraus channels,
\begin{equation}\label{eq:Kraus_2unitary}
    K_-=\sqrt{1-p}\,\check{R}(i q_-),\quad
    K_+=\sqrt{p}\,\check{R}(i q_+),
\end{equation}
which no longer satisfies the Yang-Baxter equation~(\ref{eq:YBE}). We thus arrive at the strong conclusion that even the simplest local dynamics (i.e., the convex combination of free unitaries) can lead to nonintegrable quantum circuits. This result highlights the special, and rather nontrivial, nature of the construction of the Hubbard-Kraus circuit above.
Below, we will give numerical evidence for the breaking of integrability in the preceding two examples (dubbed XXZ circuit and two-free-channel circuit).

\section{Symmetries and spectral statistics}
\label{sec:symmetries}
According to the quantum chaos conjectures~\cite{berry1977,bohigas1984} of dissipative systems~\cite{grobe1988,sa2019CSR}, the statistics of the complex eigenvalues of an integrable circuit are the same as those of uncorrelated random variables (henceforth, Poisson statistics), while nonintegrable models follow the predictions of random matrix theory (RMT), in the corresponding symmetry class. 
The comparison can only be done once all the (unitary and mutually commuting) symmetries of the model have been resolved, i.e., separating sectors with a fixed set of eigenvalues of the unitary symmetries.\footnote{Otherwise, levels from different symmetry sectors overlap without interacting and one obtains apparent Poisson statistics regardless of the actual statistics.}

Let us describe the unitary symmetries of our circuits (see Appendix~\ref{app:unitary_symmetries} for details). Because of the structure of the quantum circuit~(\ref{eq:Psi_def}), all the considered models are invariant under translation by two sites ($\comm{\Psi}{\mathbb{T}^2}=0$) and, therefore, have $L/2$ sectors of conserved quasi-momentum $k\in\{0,1,\dots,L/2-1\}$.

Furthermore, the circuits are invariant under simultaneous space translation by one site (half a unit cell) and temporal translation by one circuit layer (half a time step), which can be encoded in the commutation relation $\comm{\mathbb{T}\Phi}{\mathbb{T}^\dagger \Phi}=0$.
For a sector of fixed quasi-momentum $k$, $(\Psi)_k=e^{-4\pi i k/L}(\mathbb{T}\Phi)_k^2$, where $(A)_k\equiv\mathbb{P}_k A \mathbb{P}_k$ and $\mathbb{P}_k$ are orthogonal momentum-projection operators:
\begin{equation}\label{eq:momentum_projector}
\mathbb{P}_k=\frac{2}{L}\sum_{n=0}^{L/2-1}\mathbb{T}^{2n} \exp{-2\pi i\frac{kn}{L/2}}\,.
\end{equation}
Therefore, resolving the space-time symmetry of $\Psi$ amounts to examining the spectral statistics of $(\mathbb{T}\Phi)_k$.

Besides the kinematic symmetries of the circuit, the XX and XXZ $\check{R}$-matrices display conservation of (total) magnetization in each (bra and ket) copy of the system independently.\footnote{This also restricts the allowed incoherent processes to dephasing, which is the case in all our models.}
Accordingly, the quantum map $\Phi$ splits into $(L+1)^2$ sectors, each of dimension $N=\binom{L}{M_\uparrow}\binom{L}{M_\downarrow}$, where $M_\uparrow$ and $M_\downarrow$ denote the total magnetization in the two copies.

Once inside a fixed sector of the unitary symmetries, the symmetry class to which each circuit belongs is determined by its behavior under transposition (which can be understood as non-Hermitian time reversal) instead of complex conjugation~\cite{hamazaki2020,lieu2020}. Transposition symmetry imposes local correlations and completely determines the short-distance spectral statistics~\cite{grobe1989,hamazaki2020}. We argue in Appendix~\ref{app:transposition_symmetry} that both the Hubbard-Kraus and two-free-channel circuits admit a transposition symmetry; the latter---being nonintegrable---has, therefore, the same spectral statistics as matrices from class AI$^\dagger$~\cite{kawabata2019,hamazaki2020,sa2019CSR} (complex symmetric random matrices with Gaussian entries). In contrast, the XXZ circuit breaks transposition symmetry and has, therefore, the same spectral statistics as matrices from the Ginibre Orthogonal Ensemble (GinOE, real asymmetric random matrices with Gaussian entries).\footnote{
Because matrices from the Ginibre Orthogonal Ensemble (GinOE) and Ginibre Unitary Ensemble (GinUE) differ by their behavior under complex conjugation and not transposition, they share the same spectral correlations. For this reason, the large-$N$ results presented for the GinUE in Ref.~\cite{sa2019CSR} carry over to the GinOE.}

\begin{figure}[tbp]
	\centering
	\includegraphics[width=\columnwidth]{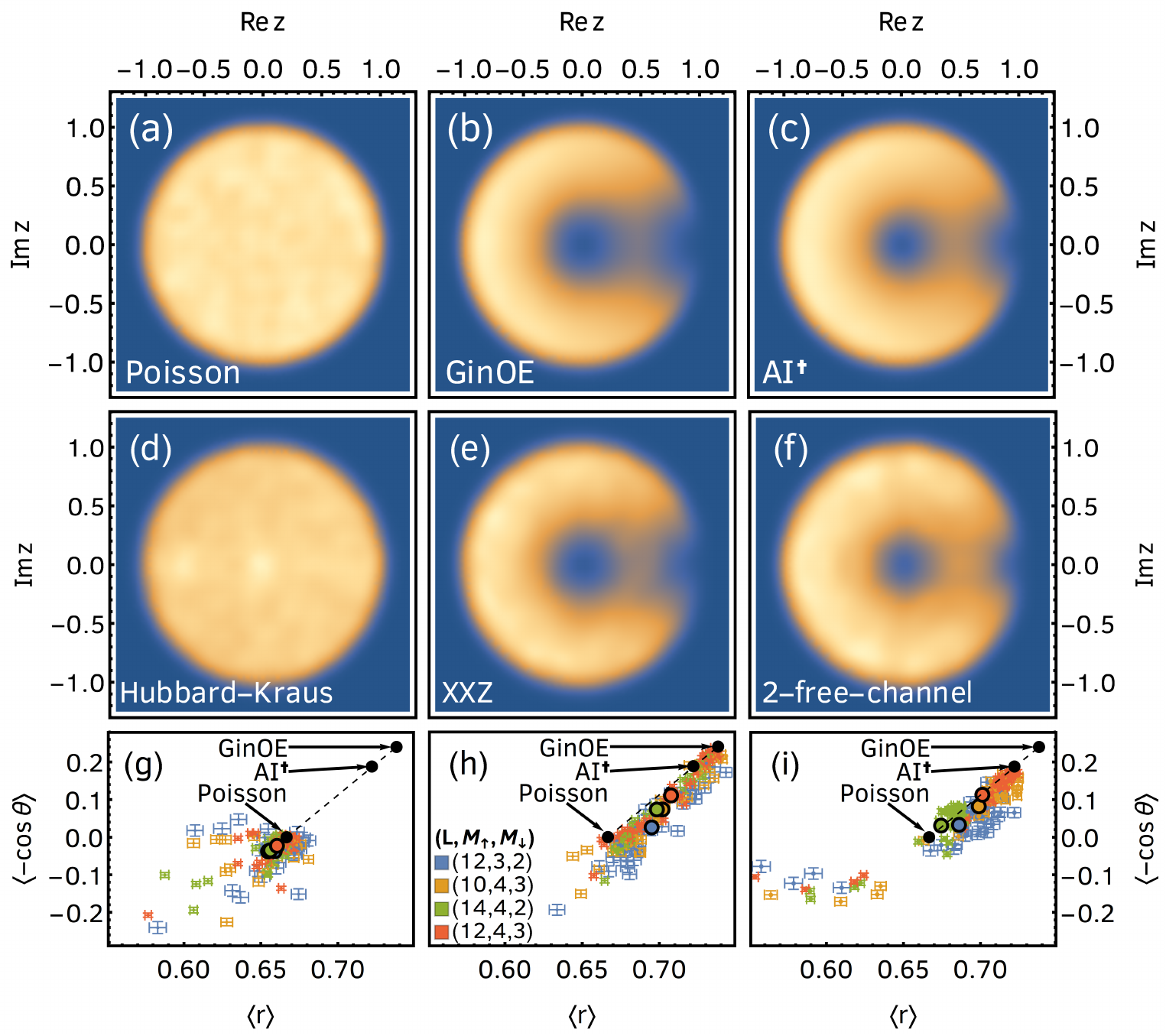}
	\caption{(a)--(c): CSR distributions obtained from sampling $10^5$ uncorrelated random variables (Poisson spectrum) (a) or exact diagonalization of $10^4\times 10^4$ random matrices from the GinOE (b) or class AI$^\dagger$ (c) [$104$ realizations superimposed in (b) and (c)]. (d)--(f): CSR distributions of (a single realization of) the operator $(\mathbb{T}\Phi)_k$ for the Hubbard-Kraus (d), XXZ (e), and two-free-channel (f) circuits. The eigenvalues were obtained by exact diagonalization in sectors of fixed $L=12$, $M_\uparrow=4$, $M_\downarrow=3$, and $k=0,1,2$---leading to three sectors of fixed quasi-momentum $k$ with $18150$ eigenvalues each, the ratios of which were then superimposed. The numerical parameters were chosen as follows: $q_+=-0.2$, $q_-=0.6$, $p=0.55$ (for all three circuits) and $\gamma_+=\gamma_-=1$ (for the XXZ circuit). (g)--(i): $\av{r}$ versus $\av{-\cos\theta}$ plots for the (g) Hubbard-Kraus, (h) XXZ, and (i) two-free-channel circuits, obtained by randomly sampling (40 samples) $p$, $q_\pm$, and $\gamma_\pm$, for different $L$, $M_\uparrow$, and $M_\downarrow$ and fixed $k=1$. Larger, black-rimmed dots mark the average (center of mass) for each symmetry sector. The points accumulate around the Poisson~$(2/3,0)$, AI$^\dagger$~$(0.722,0.188)$, and GinOE~$(0.738,0.241)$ points or spread over the line between them.}
	\label{fig:ratios}
\end{figure}

\section{Numerical results}
\label{sec:numerics}
To probe the statistics of the dissipative quantum circuits, we consider the complex spacing ratios (CSRs)~\cite{sa2019CSR} of the eigenvalues of the operator $(\mathbb{T}\Phi)_k$. We denote the set of eigenvalues by $\{\Lambda_j\}$ and, for each $\Lambda_j$, we find its nearest neighbor, $\Lambda_j^\mathrm{NN}$, and its next-to-nearest neighbor, $\Lambda_j^\mathrm{NNN}$. The CSRs are defined by $z_j=(\Lambda_j^\mathrm{NN}-\Lambda_j)/(\Lambda_j^\mathrm{NNN}-\Lambda_j)$. In the thermodynamic limit, the probability distribution of $z_j$ is flat on the unit disk for Poisson statistics; for non-Hermitian random matrices, it has a characteristic C-shape (an analytic surmise is given in Ref.~\cite{sa2019CSR}), see Figs.~\ref{fig:ratios}(a)--\ref{fig:ratios}(c). In Figs.~\ref{fig:ratios}(d)--\ref{fig:ratios}(f) we plot the CSR distribution of $(\mathbb{T}\Phi)_k$ for, respectively, the Hubbard-Kraus, XXZ, and two-free-channel circuits (eigenvalues obtained from exact diagonalization). The flat distribution of the integrable Hubbard-Kraus circuit (d) and the C-shaped distribution of the chaotic XXZ (e) and two-unitary-channel (f) circuits are clearly visible. In the latter two cases, the CSR distribution also allows us to distinguish the different symmetry classes to which the circuits belong.

To provide a more quantitative measure of spectral chaoticity, we express the CSR in polar coordinates, $z=r\exp{i\theta}$, and characterize its distribution by two numbers, namely, the mean ratio $\av{r}$, measuring the degree of radial level repulsion, and the angular correlation $\av{-\cos\theta}$. A Monte Carlo sampling over the quantum circuits with varying model parameters then yields an $\av{r}$ versus $\av{-\cos\theta}$ scatter plot, which can again be compared with the results for Poisson random variables and GinOE and AI$^\dagger$ matrices. Details on the numerical procedure can be found in Appendix~\ref{app:details_numerics}. Figures~\ref{fig:ratios}(g)--\ref{fig:ratios}(i) show the $\av{r}$ versus $\av{-\cos\theta}$ plots for the Hubbard-Kraus, XXZ, and two-free-channel circuits, respectively. For the Hubbard-Kraus circuit (g), which is integrable by construction, we obtain a high concentration of points around the Poisson point, even for modest system sizes. For the chaotic XXZ circuit (h) of the same system sizes, while points spread over the line connecting the Poisson and GinOE points, there is now a high accumulation of data around the GinOE point, signaling integrability breaking. Note that the center-of-mass values are flowing to the GinOE point as sector dimension increases. Finally, the two-free-channel circuit (i) displays the same qualitative integrability-breaking behavior as the XXZ circuit and, for the largest system sizes, has reached the AI$^\dagger$ point.

\section{Conclusions}%
\label{sec:conclusions}
Let us summarize the two key findings of this paper, addressing the critical question we posed at the start, namely, if integrability methods can be extended to the realm of dissipative quantum circuits. First, we answered it in the positive, by showing that Shastry's celebrated $\check{R}$-matrix of the Fermi-Hubbard model can be interpreted as a unital CPTP map of a pair of qubits (spins 1/2), for imaginary values of interaction and spectral parameters. By consequence of the Yang-Baxter equation, this implies integrability of the brickwork open circuit built from such nonunitary two-qubit maps.
This result opens a new avenue for studying general integrable open (driven/dissipative) quantum Floquet circuits. For example, our result straightforwardly generalizes to $\mathrm{SU}(d)$ open qudit circuits using Maassarani's $\check{R}$-matrix~\cite{maassarani1998}. 
Second, our construction shows that building integrable dissipative circuits is highly nontrivial, in the sense that deformations of the Hubbard-Kraus circuit (even convex combinations of local free-fermion unitaries) are nonintegrable.

Finally, we note that for a real interaction parameter $u\in \mathbb R$ the staggered transfer matrix (\ref{eq:monodromy_matrix}) generates a \emph{unitary} Floquet circuit (\ref{eq:Psi_TransferMatrix}) for a $2\times L$ spin ladder which represents an integrable trotterization of the Hermitian Fermi-Hubbard model by taking
$\lambda=i\tau$, $\mu=0$, where $\tau\in\mathbb R$ is the time step.
This remarkable side result parallels the result~\cite{vanicat2018} for the Heisenberg chain.

\begin{acknowledgments}
T.P.\ thanks D.\ Bernard and F.\ Essler for inspiring discussions and Institut Henri Poincar\'e for hospitality in the last stage of this work.
L.S.\ acknowledges support by FCT through Grant No.\ SFRH/BD/147477/2019. L.S.\ and P.R.\ acknowledge support by FCT through Grant No.\  UID/CTM/04540/2019. T.P.\ acknowledges ERC Advanced Grant No. 694544-OMNES and ARRS Research Program No. P1-0402.	
\end{acknowledgments}

\renewcommand{\appendixname}{APPENDIX}
\appendix

\section{\texorpdfstring{\uppercase{Kraus representation and unitary symmetries of the extensive brickwork circuit}}{Unitary symmetries}}\label{app:unitary_symmetries}%

In this Appendix, we recast the extensive circuit into the Kraus representation and use it to give a more detailed exposition of the unitary symmetries (kinematic and dynamical) of the various Kraus circuits discussed in the paper.

\subsection*{Kraus representation of the extensive circuit}%
To build the extensive quantum circuit of length $L$ out of the elementary two-site building blocks in the Kraus representation, we define a row Kraus operator $F_{\underline{\nu}}$ by tensoring $L/2$ copies of the elementary Kraus operators $K_\pm$, $F_{\underline{\nu}}=\bigotimes_{j=1}^{L/2}K_{\nu_{2j}}$.
Here, $\underline{\nu}=(\nu_2,\nu_4,\dots,\nu_L)$ is a multi-index with all two-site indices, $\nu_{2j}=\pm$, and $K_{\nu_{2j}}$ is a Kraus operator coupling sites $2j-1$ and $2j$. The quantum map corresponding to the entire row is then $\Phi=\check{\mathcal R}^{\otimes L/2}=\sum_{\underline{\nu}}F_{\underline{\nu}}\otimes F_{\underline{\nu}}^*$ (where tensor-product factors are reordered such that all second-copy degrees of freedom come after the first copy). 
The second row of the circuit is again obtained by translation by one site, $\mathbb{T}^\dagger\Phi\mathbb{T}$. Then, one complete time step is given by
\begin{align}
\label{eq:Psi_Kraus}
    \Psi
    &=\mathbb{T}^\dagger\Phi\mathbb{T} \Phi\\
\label{eq:circuit_Kraus}
    &=\sum_{\underline{\nu}}\ 
    \includegraphics[width=0.7\columnwidth,valign=c]{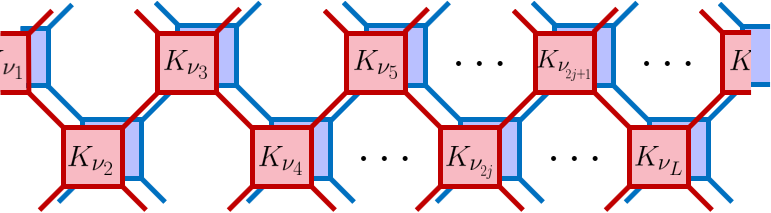}
\end{align}
where the superimposed layers represent the two copies of the system---red gates act on the ket of the density matrix while complex conjugate blue gates act on the bra.

\subsection*{Translational invariance}%
The two-site translation invariance of the circuit~(\ref{eq:circuit_Kraus}), $\comm{\Psi}{\mathbb{T}^2}=0$, leads to the conservation of quasi-momentum. Since $\mathbb{T}^L=\id$, the eigenvalues of $\mathbb{T}$ are $\exp{2\pi i(k/L)}$, with $k=0,1,\dots,L-1$. Then, the translational-invariant Kraus circuit with $L$ sites ($L$ even) has $L/2$ sectors of conserved quasi-momentum $k\in\{0,1,\dots,L/2-1\}$. Note that each row Kraus operator $F_{\underline{\nu}}$ is not translationally invariant, only their sum is, and hence there is no conservation of momentum in each bra/ket copy individually. To project states into sectors of fixed $k$, we use the orthogonal projection operators~(\ref{eq:momentum_projector}).

\subsection*{Magnetization conservation}%
The XX and XXZ $\check{R}$-matrices display conservation of (total) magnetization (or particle-number in a fermion picture) in each copy of the system independently. In these conditions, for a given copy of the system, each row Kraus operator $F_{\underline{\nu}}$ splits into $L+1$ sectors of total magnetization $S_z=M$, where $S_z$ acts on the computational-basis states as $S_z\ket{s_1,\dots,s_L}=\sum_{j=1}^Ls_j\ket{s_1,\dots,s_L}$. Each sector $M$ has dimension $\binom{L}{M}$. Accordingly, the quantum map $\Phi$ splits into $(L+1)^2$ sectors, each of dimension $N=\binom{L}{M_\uparrow}\binom{L}{M_\downarrow}$ (here $M_\uparrow$ and $M_\downarrow$ denote the magnetization in the two copies).
We restrict ourselves to sectors with $M_{\uparrow,\downarrow}\neq L/2$ to avoid an additional $\mathbb{Z}_2$ spin-flip (particle-hole) symmetry. Finally, there is another $\mathbb{Z}_2$ symmetry connecting the two copies of the system; we also avoid this symmetry by considering only sectors with $M_\uparrow\neq M_\downarrow$.

\section{\texorpdfstring{\uppercase{Transposition-symmetry classes}}{Transposition-symmetry classes}}\label{app:transposition_symmetry}%

In this Appendix, we elaborate on the symmetry classification of general non-Hermitian matrices and CPTP generators, in particular in terms of the transposition symmetry. We also argue for the presence (absence) of transposition symmetry in the Hubbard-Kraus and two-free-channel (XXZ) circuits. 

\subsection*{Symmetry classification of non-Hermitian matrices and CPTP generators}

The symmetry class to which each circuit belongs is determined by its antiunitary (and anticommuting unitary) symmetries. While there are 38 symmetry classes of non-Hermitian matrices~\cite{bernard2002,kawabata2019,zhou2019}---dictated by the behavior under sign inversion, complex conjugation, transposition, and Hermitian conjugation---considering only the generators of CPTP dynamics restricts the allowed symmetry classes back to ten~\cite{lieu2020,altland2020}, which are in one-to-one correspondence with the Altland-Zirnbauer~\cite{altland1997,chiu2016} classes of closed quantum systems. Indeed, for a genuine quantum channel $\Psi$, its Hermiticity-preserving property guarantees the existence of a symmetry $\Psi=S\Psi^* S^\dagger$ for some unitary $S$, while complete positivity forbids the existence of a symmetry $\Psi=-S \Psi^\top S^\dagger$ or $\Psi=-S\Psi S^\dagger$. Of the remaining three types of symmetries, only transposition symmetry (i.e., the existence of a unitary $T$ such that $\Psi=T \Psi^\top T^\dagger$ with $TT^*=\pm1$) imposes local correlations and, hence, completely determines the short-distance spectral statistics~\cite{grobe1989,hamazaki2020}. So, while there are ten remaining symmetry classes, only three different universal statistics exist, differing by the amount of level repulsion. In the absence of transposition symmetry, the generator is represented by a general real asymmetric matrix and shares spectral statistics with random matrices from the GinOE. If there is a unitary $T$ satisfying $TT^*=+1$, then the generator shares the spectral statistics with the complex symmetric matrices from class AI$^\dagger$.

\subsection*{Transposition symmetry of the Kraus circuits}

\begin{figure*}[t]
    \centering
    \includegraphics[width=\textwidth]{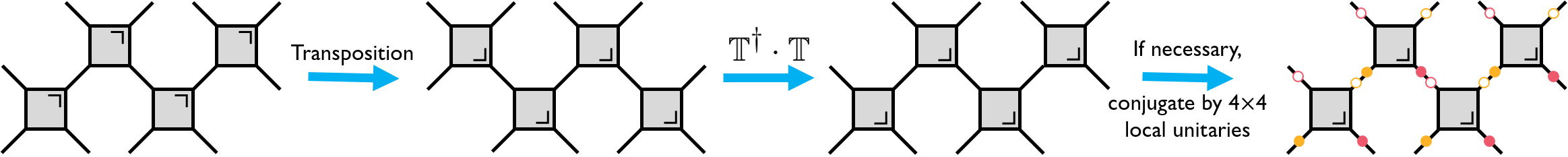}
    \caption{Schematic representation of the steps involved in determining whether there exists a transposition symmetry of the circuit. Gray gates represent local quantum maps $\phi$, while orange and magenta filled circles depict local unitaries $v_-,v_+\in \mathrm{SU}(4)$, respectively, and empty circles their inverses. Transposition is signaled by the flip of the wedge in the corner of the local maps. To respect the kinematical symmetries of the circuit, the allowed unitary transformations are one-site translations and the local unitaries $v_\pm$.}
    \label{fig:SM_transposition_circuit}
\end{figure*}

We now analyze the behavior of the three circuits considered in the paper under transposition. We start by showing that it suffices to consider the properties of elementary two-site quantum maps. Taking the transpose of the circuit we find
$\Psi^\top=\Phi^\top \mathbb{T}^\top \Phi^\top \mathbb{T}^*=\Phi^\top \mathbb{T}^\dagger \Phi^\top \mathbb{T}$. We want to bring it back to $\Psi$ by a unitary transformation that satisfies the symmetries of the model. These include translations (necessary to bring the circuit back to the correct order of applying first gates as odd-even bonds followed by even-odd bonds) and, possibly, local $4\times 4$ unitaries in each local Hilbert space (local gauge transformations). The procedure is depicted pictorially in Fig.~\ref{fig:SM_transposition_circuit}.

We conclude that the circuit satisfies the transposition symmetry $\Psi=T\Psi^\top T^\dagger$ if the local quantum maps $\phi$ satisfy
\begin{equation}\label{eq:transposition_condition}
    \phi=
    \includegraphics[width=0.15\textwidth,valign=c]{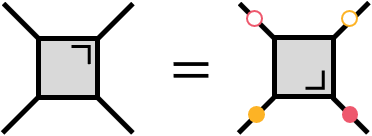}
    = (v_-\otimes v_+)\phi^\top (v_+^\dagger \otimes v_-^\dagger),
\end{equation}
for some $v_-,v_+\in \mathrm{SU}(4)$ respecting the dynamical symmetries of the circuits (i.e., magnetization conservation). Moreover, from Eq.~(\ref{eq:Kraus_rep}), we see that the behavior under transposition of $\phi$ is fully determined by the behavior of the Kraus operators $K_\pm$ of each circuit (since tensoring and transposing commute). It thus suffices to analyze the behavior of the Kraus operators under transposition.

\subsubsection*{Two-free-channel circuit}%
The local quantum map $\phi$ is a convex combination of a pair of unitary channels, whose Kraus operators are given in Eq.~(\ref{eq:Kraus_2unitary}). The transposition symmetry is evident from Eq.~(\ref{eq:Rch_general6v}) after a unitary change of basis  by conjugation with $V=\mathrm{diag}(1,\exp\{i\pi/4\},\exp\{-i\pi/4\},1)$---which corresponds to the choice of gauge $v_-=v_+^*=V$ in Eq.~(\ref{eq:transposition_condition}). Writing out the Kraus operators explicitly (we omit all zero entries) in the new basis,
\begin{equation}
    K_\pm(\lambda_\pm)=\begin{pmatrix}
    \cos\lambda_\pm & & & \\
    & 1 & \sin\lambda_\pm & \\
    & \sin\lambda_\pm & 1 & \\
    & & & \cos\lambda_\pm
    \end{pmatrix},
\end{equation}
we see that they are complex symmetric, where we defined the spectral parameters $\lambda_\pm \equiv\lambda\pm\mu\in i\mathbb{R}$ (related to the hopping amplitudes $q_\pm$ by multiplication by $i$). (The normalization of the $\check{R}$-matrix is irrelevant for the purpose of this appendix and will be dropped throughout.) It follows that, in this basis, $\phi(\lambda_-,\lambda_+)=\phi^\top(\lambda_-,\lambda_+)$, and the circuit enjoys a transposition symmetry.

\begin{widetext}
\subsubsection*{Hubbard-Kraus circuit}%
Although the Hubbard-Kraus circuit is integrable and, therefore, exhibits Poisson spectral statistics, it is instructive to determine its symmetry class to provide contrast to the XXZ circuit case discussed below. The Kraus operators (\ref{eq:Kraus_2channel}) read as
\begin{equation}
    K_-(\lambda_-)=\begin{pmatrix}
        \cos\lambda_- & & & \\
        & 1 & -i\sin\lambda_- & \\
        & i\sin\lambda_- & 1 & \\
        & & & \cos\lambda_-
    \end{pmatrix}
    \quad \text{and} \quad
    K_+(\lambda_+)=\begin{pmatrix}
        \cos\lambda_+ & & & \\
        & 1 & i\sin\lambda_+ & \\
        & i\sin\lambda_+ & -1 & \\
        & & & -\cos\lambda_+
    \end{pmatrix}.
\end{equation}
While both Kraus operators cannot be symmetrized simultaneously by a change of basis as before, they still satisfy
\begin{equation}\label{eq:weak_transposition}
    K_-^\top(\lambda_-)=K_-(-\lambda_-)=K_-(\lambda_-^*)
    \quad \text{and} \quad 
    K_+^\top(\lambda_+)=K_+(\lambda_+).
\end{equation}
It follows that $\phi(\lambda_-,\lambda_+)=\phi^\top(\lambda_-^*,\lambda_+)$. For nonintegrable cases, generalizing the local Hubbard-Kraus map to more than two Kraus operators of the form~(\ref{eq:Kraus_2channel}), we have made the empirical observation that the equality of the Kraus operators and their transposes \emph{up to complex conjugation of the imaginary spectral parameters} is enough to guarantee the convergence of their spectral statistics to those of the AI$^\dagger$ class.
This condition is fulfilled by the Hubbard-Kraus map.

\subsubsection*{XXZ circuit}%
By introducing interactions into the coherent dynamics, i.e., by considering Kraus operators
\begin{equation}
\begin{split}
    K_-(\lambda_-)&=\begin{pmatrix}
        \sin(\gamma_-+\lambda_-) & & & \\
        & \sin\gamma_- & -i\sin\lambda_- & \\
        & i\sin\lambda_- & \sin\gamma_- & \\
        & & & \sin(\gamma_-+\lambda_-)
    \end{pmatrix}
    \quad \text{and}\\
    K_+(\lambda_+)&=\begin{pmatrix}
       \sin(\gamma_++\lambda_+) & & & \\
        & \sin\gamma_+ & i\sin\lambda_+ & \\
        & i\sin\lambda_+ & -\sin\gamma_+ & \\
        & & & -\sin(\gamma_++\lambda_+)
    \end{pmatrix},
\end{split}
\end{equation}
it follows that neither can both Kraus operators be simultaneously symmetrized nor do they satisfy Eq.~(\ref{eq:weak_transposition}). Hence, the XXZ circuit does not enjoy a transposition symmetry.
\end{widetext}

\section{\texorpdfstring{\uppercase{Details on the numerical analysis}}{Details on the numerical analysis}}
\label{app:details_numerics}%
In this Appendix, we discuss in more detail the random sampling of the circuits and the numerical analysis of CSRs.

To obtain the scatter plots of Figs.~\ref{fig:ratios}(g)--\ref{fig:ratios}(i), we express the CSR in polar coordinates and characterize its distributions with two numbers (the mean ratio $\av{r}$ and the angular correlation $\av{-\cos\theta}$).

For each model, we randomly sample the two independent hopping parameters $q_\pm$ from a standard normal distribution (i.e., zero mean, unit variance), the channels' relative weight $p$ from a uniform distribution on $[0,1]$, and, in the case of the XXZ circuit, the anisotropy parameters $\gamma_\pm$ from a uniform distribution on $(-\pi,\pi]$. For a fixed random realization of the parameters, we exactly diagonalize the quantum circuit for four different system sizes and conserved magnetization sectors---$(L,M_\uparrow,M_\downarrow)$=$(12,3,2)$, $(10,4,3)$, $(14,4,2)$, and $(12,4,2)$, corresponding to sector sizes $N$=$2420$, $5040$, $13013$ and $18150$, respectively---and compute the pairs ($\av{r},\av{-\cos\theta}$). A Monte Carlo sampling of the quantum circuits then yields a $\av{r}$ versus $\av{-\cos\theta}$ scatter plot.

There are three special points: For a set of uncorrelated (Poisson) random variables we have exactly $(\av{r},\av{-\cos\theta})=(2/3,0)$, while for random matrices from the GinOE and class AI$^\dagger$ we have $(\av{r},\av{-\cos\theta})\approx(0.738,0.241)$ and $(\av{r},\av{-\cos\theta})\approx(0.722,0.188)$, respectively~\cite{sa2019CSR}. When approaching the thermodynamic limit, we expect that the points obtained by sampling an integrable circuit concentrate around the Poisson point, while they accumulate around the RMT point of the respective symmetry class when sampling from a nonintegrable model. For finite system sizes $L$, the points spread over the line connecting the two fixed points and one tries to determine to which one of them the scatter points are flowing as $L$ increases. This last step may be difficult to perform if only small system sizes are available and/or finite-size effects are pronounced. Moreover, the flow may be nonmonotonic; for instance, it may depend more strongly on the system length $L$ than on the magnetization sectors $M$, or sectors with different parity of $M$ may flow differently.

Finally, there are deviations from the pattern described above if the spectrum is close to one-dimensional, for instance, if the quantum map is very close to being unitary (which happens when $p$ is very close to either $0$ or $1$). For this reason, we have only considered quantum circuits for which $\abs{z_\mathrm{max}}-\abs{z_\mathrm{min}}>0.2$, where $z_{\mathrm{\max}(\mathrm{min})}$ are the eigenvalues of $\mathbb{T}\Phi$---in the the appropriate symmetry sector---with largest (smallest) absolute value.

\bibliography{bibfile}

\end{document}